\documentclass[aps,prl,twocolumn,letterpaper,superscriptaddress,showpacs]{revtex4}
\usepackage{keyval}%
\usepackage{graphicx}
\usepackage{dcolumn}
\usepackage{bm}
\usepackage{color}

\setkeys{Gin}{width=0.8\columnwidth,clip=true}

\newcommand{\ignore}[1]{}

\begin{document}
\title{Ferro-Orbital Order and Strong Magnetic Anisotropy in the Parent Compounds of
Iron-Pnictide Superconductors}
\author{Chi-Cheng Lee}
\author{Wei-Guo Yin}
\affiliation{Condensed Matter Physics and Materials Science
Department, Brookhaven National Laboratory, Upton, New York 11973,
USA}
\author{Wei Ku}
\affiliation{Condensed Matter Physics and Materials Science
Department, Brookhaven National Laboratory, Upton, New York 11973,
USA}%
\affiliation{Physics Department, State University of New York, Stony
Brook, New York 11790, USA}

\date{\today}

\begin{abstract}
The puzzling nature of magnetic and lattice phase transitions of
iron pnictides is investigated via a first-principles Wannier
function analysis of representative parent compound LaOFeAs. A rare
ferro-orbital ordering is found to give rise to the recently
observed highly anisotropic magnetic coupling, and drive the phase
transitions---without resorting to widely employed frustration or
nesting picture. The revealed necessity of the additional orbital
physics leads to a correlated electronic structure fundamentally
distinct from that of the cuprates. In particular, the strong
coupling to the magnons advocates active roles of light orbitons in
spin dynamics and electron pairing in iron pnictides.
\end{abstract}


\maketitle

With a striking similarity to the cuprate families~\cite{Lee},
recently discovered high-temperature superconductivity in
iron pnictides emerges upon suppression of magnetic order via
additional charge doping~\cite{Kamihara,Cruz}. This new example of
close proximity of superconductivity to the magnetic phase
has raised again the long standing questions concerning the intimate
relationship between these two seemingly different (or even
exclusive) phases of
solids~\cite{Yin,Zhao,Mazin,Wang,Dong,Raghu,Kuroki,Moreo,Si,Fang,Yildirim,Han}.
Following the cuprate research, a big thrust of current efforts has
been to establish superconductivity in these compounds via magnetic
correlations,
despite the apparently diverse perspectives of the magnetism
itself~\cite{Yin,Zhao,Mazin}.
Clearly, a solid understanding of the parent compounds and their
magnetism is an essential first step toward a convincing resolution
of superconductivity in the doped systems, especially within the
heavily discussed spin fluctuation scenario of
pairing~\cite{Yin,Zhao,Mazin,Wang,Dong,Raghu,Kuroki,Moreo,Si,Fang,Yildirim,Han}.

The parent undoped compounds of the iron pnictides have a quite
unusual in-plane magnetic structure. Unlike the antiferromagnetic
(AF) magnetic moments along both the $x$ and $y$ directions in the
copper oxides (the G-AF structure), the magnetic moments in the iron
pnictides are only AF in the $x$ direction, but align
ferromagnetically along the $y$ direction. In the local picture,
this curious stripe-like (or C-AF) structure is currently explained
~\cite{Si,Fang,Yildirim} by requiring a strong next nearest neighbor
(NNN) Heisenberg AF coupling, $J_{2}$, larger than half of the
nearest neighbor (NN) AF coupling, $J_{1}$, such that the G-AF
structure favored by $J_{1}$ is suppressed energetically. The
competition between C-AF and G-AF implies a strong magnetic
frustration~\cite{Si,Fang,Yildirim}. This frustration was argued to
account for the observed small iron spin moment $\sim
0.36$~$\mu_\mathrm{B}$~\cite{Cruz} and to promote superconducting
order via relief of the magnetic entropy~\cite{Si}. In addition, a
fluctuating electron nematic order was predicted in the spin
frustration case~\cite{Fang} to account for the structural
transition at slightly higher temperature~\cite{Cruz}, and was used
to support the close relation between the physics of the cuprate and
the iron-pnictide superconductors~\cite{Fang}.

However, if the system is really that frustrated, the rather high
magnetic transition temperature
($T_\mathrm{N}\sim$137~K)~\cite{Cruz} would be difficult to
understand. Furthermore, against the overall symmetry of the system,
a surprisingly strong anisotropic NN coupling in the $x$ and $y$
direction was identified very recently from the inelastic neutron
scattering measurements on CaFe$_{2}$As$_{2}$~\cite{Zhao},
consistent with results of recent DFT calculations~\cite{Yin,Han}.
This enormous anisotropy suggests that a \textit{strong} rotational
symmetry breaking has taken place prior to the magnetic ordering,
setting a very stringent constraint to the correct microscopic
understanding of the magnetism. These results have been considered
as direct evidence against~\cite{Mazin} the above local Heisenberg
picture and against~\cite{Zhao} the alternative spin-density-wave
picture for itinerant electrons with ``nesting'' Fermi
surface~\cite{Mazin,Wang,Dong,Raghu,Kuroki}. Clearly, a
comprehensive new picture is urgently needed to explain the
microscopic origin of the magnetic structure together with the
observed strong anisotropy, and in particular to address the potential
additional symmetry breaking, considering its profound implications to the
electronic structure and to the superconductivity.

In this Letter, we report a first-principles Wannier function
analysis~\cite{Ku,Weiguo} of the electronic structure of the representative parent
compound, LaOFeAs. A purely electronic ferro-orbital
order is found to spontaneously break the rotational symmetry and
drive the observed magnetic and structural transitions without
resorting to the widely employed Fermi surface nesting or magnetic
frustration. In great contrast to the cuprates, our study reveals
the essential roles of the orbital degree of freedom (especially its
short-range correlation) in the
electronic structure of iron pnictides. In particular, the expected
light mass of orbiton and its
strong coupling to the magnon advocate its active roles in magnon
decay and electron pairing.

The electronic structure of LaOFeAs is calculated within local spin
density approximation (LSDA) and its local interaction extension
(LDA+$U)$ of density functional theory, implemented via full
potential, all-electron, linearized augmented plane wave
basis~\cite{Schwarz}. The atomic positions are relaxed at the
experimental lattice constants of the undistorted lattice at 175
K~\cite{Cruz}. A unit cell containing four Fe atoms in the plane is
adopted to accommodate the observed C-AF structure.
The $x$ axis is
chosen along the AF ordered direction of C-AF.
A set of energy-resolved symmetry-respecting Wannier functions (WFs)
is then constructed~\cite{Ku,Weiguo}
that spans the
complete Hilbert space within 3eV of the Fermi level.

In significant contrast to the maximally localized WF method~\cite{Marzari}
used in the repvious studies~\cite{Kuroki,Cao},
our construction exploits the ``gauge freedom''~\cite{Marzari} of the WFs to
achieve localization within constraints of the point-group symmetry~\cite{Ku,Weiguo}.
Specifically, the cores of our Fe $d_{xz}$ and $d_{yz}$ WFs extend toward the
directions of the magnetic ordering, not the As atoms.
Consequently, the one-particle density matrix, $\rho_{ij}=<$WF$_i|\hat{\rho}|$WF$_j>$,
is locally diagonalized automatically, allowing direct detection of $spontaneous$
orbital polarization~\cite{supplement}.

\begin{figure}[tbp]
\includegraphics[width=0.6\columnwidth,clip=true,angle=270]{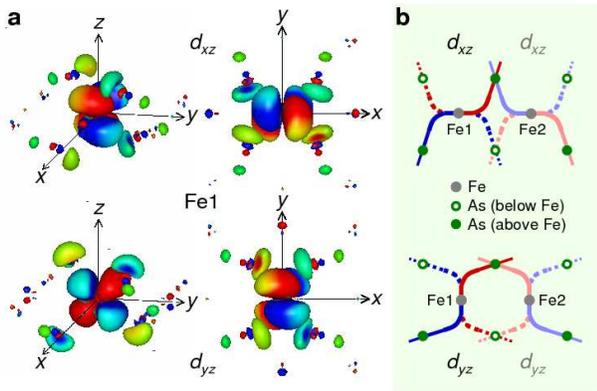}
\caption{\label{fig:fig2}
(color online). \textbf{a}, Side
view (left) and top view (right) for the Fe1 3$d_{xz}$ (upper panels) and
3$d_{yz}$ (lower panels) Wannier orbitals obtained from nonmagnetic calculations,
colored by their positive (red) and negative (blue) gradient. \textbf{b}, Sketches of top view of these orbitals
on two nearest neighboring Fe atoms along the $x$ direction (Fe1 and Fe2).
The solid (dashed) lines denote the tails above (below) the Fe
plane.}
\end{figure}

Illustrated in Fig.~\ref{fig:fig2}a are two of the resulting WFs
most relevant to our further analysis (Fe 3$d_{xz}$ and 3$d_{yz})$,
as they are the only $d$-orbitals anisotropic in the $xy$
directions. To simplify the visualization, a sketch illustration of
these WFs at the two neighboring sites is given in
Fig.~\ref{fig:fig2}b. The dramatic effects of hybridizing with
tetrahedral positioned As 4$p$ orbitals can be clearly observed from
the anti-bonding ``tails'' of the WFs. Notice, in particular, how
such hybridization bends the tails of the WFs \textit{perpendicular}
to their original directions, and significantly modifies the local
point group symmetry. In great contrast to the normal
two-dimensional character, the $d_{xz}$ and $d_{yz}$ WFs of LaOFeAs
present a clasp-like shape around the iron center. Serious
consequence of this hybridization will be discussed in detail below.

\begin{figure}[tbp]
\includegraphics[width=0.4\columnwidth,clip=true,angle=270]{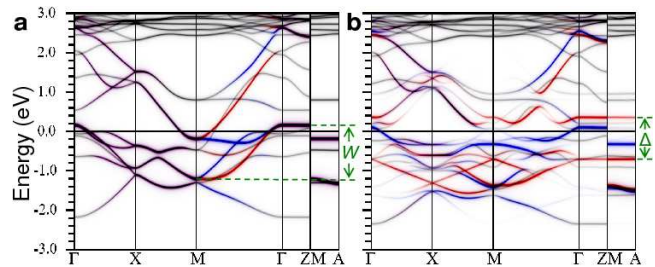}
\caption{\label{fig:fig3} (color online). Electronic band structures of
(\textbf{a}) nonmagnetic and (\textbf{b}) C-AF configurations,
represented in the nonmagnetic Brillouin zone. The weight of the Wannier Fe
3$d_{xz}$ and 3$d_{yz}$ orbitals are presented via blue and red colors,
respectively.}
\end{figure}

With the help of these WFs, the LSDA band structures for non-magnetic (NM)
and C-AF LaOFeAs are compared in Fig.~\ref{fig:fig3}.
To explicitly illustrate the effects of broken periodicity,
the band structures are presented in the reciprocal space of the nominal NM
unit cell containing only two Fe atoms. In this representation, additional
gap openings and ``shadow bands'' can be clearly observed in the C-AF case,
whose intensity reflects the strength of these bands' coupling to the magnetic
order.

Notice that the bands near the Fermi level (zero energy)
primarily consist of the above mentioned Fe $d_{xz}$ and
$d_{yz}$ WFs (c.f.: Fig.~\ref{fig:fig2}). In the NM case, these two
orbitals are degenerate as guaranteed by the point-group symmetry.
In the C-AF case, on the other hand, a very large splitting in the
broad vicinity of the Fermi level is observed involving the Fe
$d_{yz}$ orbital, while the Fe $d_{xz}$ orbital is only weakly
affected. This indicates that \textit{only one of the two orbitals
is significantly involved in the magnetic coupling in the C-AF
configuration, and the symmetry between the orbitals is broken.}
Indeed, as the magnetic order develops, the spin polarization
(obtained from $\rho_{ij}$) is
found to be much stronger in the Fe $d_{yz}$ orbital ($\sim 0.34
\mu_\mathrm{B}$) than in the $d_{xz}$ orbital ($\sim $0.15
$\mu_\mathrm{B}$), due to its loss of electron occupation in the
spin minority channel. Introduction of a realistic moderate local
interaction ($U $= 2 eV and Hund's coupling $J $= 0.5 eV) further
enhances this effect and increases the Fe $d_{yz}$ polarization to
0.58 $\mu_\mathrm{B}$, while the moment in the Fe $d_{xz}$ remains
small (0.23 $\mu_\mathrm{B}$) with both spin channels heavily
populated.

In addition, \textit{the gap opening, $\Delta $, of the $d_{yz}$
orbital is found to be comparable to the band width, $W$}, as shown
in Fig.~\ref{fig:fig3}. Such a large gap is commonly encountered in
strongly correlated systems where the magnetism is more conveniently
described by interacting local moments, rather than itinerant
electrons with nested Fermi surface. Therefore, we will proceed
below to build a local correlated picture for the electronic
structure.

Further insights into the microscopic local processes are revealed
by transforming the DFT Hamiltonian of the NM case to the Wannier
basis, as given in Table~\ref{tab:table1}. Unexpectedly, the leading
hopping paths of Fe $d_{yz}$ orbital are to the neighboring $d_{yz}$
and $d_{x^{2}-y^{2}}$ orbitals along the $x$ direction,
\textit{perpendicular to its original direction}. This is
\textit{anti-intuitive} since within the simple cubic symmetry the
former would have been very weak ``$\pi $-bond''-like, while the
latter would have been symmetry forbidden and identical to zero. By
contrast, the supposedly stronger ``$\sigma $-bond''-like NN
$d_{xz}-d_{xz}$ hopping along the $x$ direction is remarkably weak.
These features are qualitatively different from those used in the
previous studies~\cite{Kruger,Calderon}. Our distinctly different
results originate mathematically from the change of direction in the
WFs' hybridization tails, as visualized above in
Fig.~\ref{fig:fig2}~\cite{supplement}.
Physically, this reflects the dramatic
influence of the tetrahedral positioning of As 4$p$ orbitals on the
Fe 3$d$ orbitals, and reveals the importance
of As atom positions and Fe-As phonon modes in the electronic
structure in general.

\begin{table}[t]
\caption{\label{tab:table1}  Onsite energy (first row) and hopping
integrals among Fe 3$d$ Wannier orbitals for the nonmagnetic case
(in eV). Fe2 and Fe3 are the NN and NNN of Fe1 (c.f.
Fig.~\ref{fig:fig4}).}
\begin{ruledtabular}
\begin{tabular}{cccccc}
$\langle$WFs$ | H | $WFs$\rangle$ & Fe1 $z^2$& $x^2$-$y^2$ & $yz$ & $xz$ & $xy$ \\
\hline
Fe1 $\epsilon$ - $\mu$ & $-0.03$ & $-0.20$ & 0.10 & 0.10 & 0.34 \\
\hline
Fe2 $z^2$  & 0.13 & 0.31 & $-0.10$ & 0.00 & 0.00 \\
$x^2$-$y^2$& 0.31 &$-0.32$ & $\mathbf{0.42}$ & 0.00 & 0.00 \\
$yz$       &$-0.10$ & 0.42 & $\mathbf{-0.40}$& 0.00& 0.00 \\
$xz$       & 0.00 & 0.00 & 0.00 & $\mathbf{-0.13}$ & $-0.23$ \\
$xy$       & 0.00 & 0.00 & 0.00 & $-0.23$ & $-0.30$ \\
\hline
Fe3 $z^2$& 0.06& 0.00& $-0.08$ & 0.08& 0.26 \\
$x^2$-$y^2$& 0.00& $-0.10$ & 0.12& 0.12& 0.00 \\
$yz$ & 0.08& $-0.12$& $\mathbf{0.25}$ & $-0.07$& $-0.05$ \\
$xz$& $-0.08$& $-0.12$& $-0.07$& 0.25& 0.05 \\
$xy$& 0.26& 0.00& 0.05& $-0.05$& 0.16 \\
\end{tabular}
\end{ruledtabular}
\end{table}

\begin{figure}[t]
\includegraphics[width=0.55\columnwidth,clip=true,angle=270]{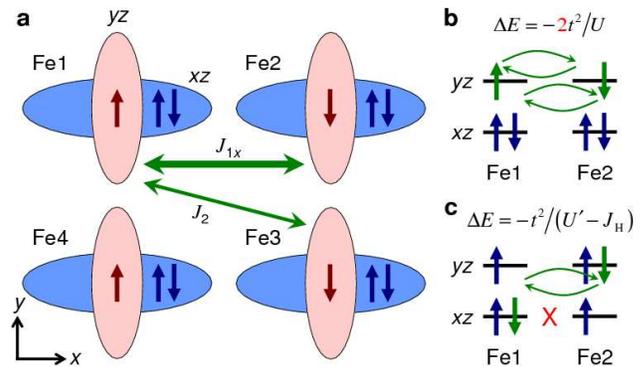}
\caption{\label{fig:fig4}
(color online). \textbf{a} Schematic of C-AF magnetic structure with highly
anisotropic NN coupling due to orbital ordering.
\textbf{b}, \textbf{c} The kinetic energy $\Delta E$
in the ferro-orbital and staggered orbital structures.
$t$ and $U$ denote the hopping parameter and
intra-orbital Coulomb repulsion, while $U'$ and $J_\mathrm{H}$ denote the inter-orbital
repulsion and Hund's exchange, respectively.}
\end{figure}

A simple ``minimal'' picture of the low-energy physics now emerges
from the above analysis that elucidates the nature of C-AF magnetic
structure together with the observed large anisotropy. Let's focus
only on the Fe $d_{xz}$ and $d_{yz}$ orbitals, as they are the only
$d$-orbitals anisotropic in the $xy$ directions. As illustrated in
Fig.~\ref{fig:fig4}, given almost doubly-occupied Fe $d_{xz}$
orbitals and almost singly-occupied, spin-polarized Fe $d_{yz}$
orbitals, the $d_{yz}$ orbitals prefer AF alignments along the
directions of the efficient hopping, to benefit from the kinetic
energy (the ``super exchange'').
Within the over-simplified strong coupling limit ~\cite{supplement}
to the second order in the hopping parameters in Table~\ref{tab:table1},
the leading AF magnetic couplings among the $d_{yz}$ subspace are the NN
coupling along the $x$ direction, $J_{1x}$, and the NNN coupling,
$J_{2}$ $\sim $ 0.4 $J_{1x}$. In comparision, the NN coupling along the
$y$ direction, $J_{1y} \sim $ 0.1 $J_{1x}$, is insignificant. This large anisotropy
is in good agreement with the current experimental~\cite{Zhao} and
theoretical observations~\cite{Han}. Clearly, \textit{the
anisotropy} \textit{owes its origin to the orbital degree of
freedom}, as the rotational symmetry breaking takes place via
orbital polarization.
Therefore, modeling the magnetic structure with
a standard Heisenberg model~\cite{Si,Fang,Yildirim,Yin,Han,Zhao}
would suffer from its very limited applicability~\cite{Yaresko}, as
it lacks flexibility to break the rotational symmetry spontaneously,
or to adjust the magnetic coupling strength according to the orbital structure.

As shown in Fig.~\ref{fig:fig4}, following these two leading AF
couplings $J_{1x}$ and $J_{2}$, the observed C-AF structure is
naturally established \textit{locally}, without resorting to Fermi
surface nesting. The seemingly ferromagnetic allignment along
the $y$ axis results primarily from the NN and NNN AF alignment
across the columns. Contrary to previous theoretical
explanation~\cite{Si,Fang,Yildirim}, in our picture the observed
C-AF structure is \textit{not} frustrated or competing with any
other magnetic structure ($e.g.$: G-AF), and thus can sustain a high
transition temperature~\cite{Cruz}.

In order to maximize the kinetic energy gain via the
super-exchange processes, the orbitals not only have to be
polarized, but also need to be ordered.
(Other magnetic/orbital configurations were found to have higher energies.)
Indeed, in the above
picture, all the sites are polarized the same way with the $d_{yz}$
orbital being less occupied and more spin polarized. This can be
considered an example of ``ferro-orbital order''~\cite{Kugel,Kruger}.
The formation of this rare orbital
order can be understood by noting again the anti-intuitive NN
hopping path along the $x$ direction (c.f. Table~\ref{tab:table1}),
dominated by only hopping between $d_{yz}$ orbitals without
$d_{xz}$-$d_{yz}$ cross hopping. As illustrated in
Fig.~\ref{fig:fig4}b, the best way to utilize the kinetic energy in
this case is indeed the ferro-orbital, AF magnetic alignment, since
one electron from \textit{both} sites benefit from the kinetic
energy. In comparison, the more common staggered-orbital,
ferromagnetic alignment (\textit{e.g.} in undoped manganites)
can utilize efficient hoppings in \textit{only one} channel
(c.f. Fig.~\ref{fig:fig4}c), despite the
additional benefit from the intra-atomic interactions~\cite{supplement}.
That is, the unique hopping path leads to a rare ground state of the undoped
iron pnictides consisting of coexisting, cooperative ferro-orbital and C-AF orders.

The ferro-orbital phase is in fact directly responsible for the
tetragonal-to-orthorhombic lattice transition observed at $T_{s}
\sim $155 K~\cite{Cruz}. As soon as long-range ferro-orbital order
takes place at $T_{OO}$($=T_s$), $d_{xz}$ orbitals becomes more
occupied macroscopically (c.f. Fig.~\ref{fig:fig4}), leading to a
longer bond in the $x$ direction, in agreement with the
experiment~\cite{Huang}. In general, it is very rare to have AF bond
longer than the ferromagnetic one in the late transition metal
compounds, since the AF super exchange grows on almost half-filled
orbitals (with less charge). In iron pnictides, this unusual long bond
is realized only because of the perpendicular extension of the
hybridization tails discussed above. This subtle behavior
distinguishes our result from a previous study~\cite{Kruger}, in
which the AF bond is less populated and thus would be shorter.

It is important to further clarify the weak coupling strength
between the elastic lattice distortion and the orbital order. When
the lattice is set to frustrate the orbital order in our calculation
by exchanging the lattice constants in the $x$ and $y$ directions,
only less than 10 meV cost per Fe is found, an order of magnitude
smaller than the super exchange energy. In essence, the orbital
order originates almost entirely from the electronic energy, and the
lattice simply follows the orbital order. On the other hand, such a
weak coupling should allow a strong short-range orbital correlation
even above $T_{s}$, offering a natural explanation to the large
anomalous signal extending to 40K above $T_{s}$ in the thermal
expansion measurement~\cite{LWang}.

Since it is the same kinetic energy gain that drives both the
magnetic and orbital orders, they are thus strongly coupled to each
other, as in the Kugel-Khomskii model~\cite{Kugel}. This is the
natural reason for the close proximity of observed $T_\mathrm{N}$
and $T_{s}$. Unlike the manganites, where the orbital order is
further stabilized by the large lattice-orbital coupling $\sim $0.9
eV~\cite{Weiguo}, the weak coupling to the elastic lattice mode here
does not help raising $T_{s}$ noticeably from $T_\mathrm{N}$. This
suggests a light effective mass of the orbiton, comparable to the
magnon mass. Thus, an intimate relationship between the excitations
in the spin and orbital channel is expected. Indeed, an efficient
orbiton-assisted decay offers a very interesting possible
explanation to the observed large line width of magnon at large
momentum~\cite{Zhao}. Furthermore, the controlling role of Fe-As-Fe
positioning in the unconventional hopping implies large isotope
effects in both magnetic~\cite{Liu} and orbital transitions.

Our study also has several implications on the high-temperature
superconductivity of doped iron pnictides. In great contrast to the
cuprates, our results reveal the essential roles of the orbital
degree of freedom in iron pnictides.
Upon doping, the long-range magnetic and orbital orders would naturally
perish rapidly through the disruption of the above superexchange process
by doping-induced charge fluctuation.
(Recall that the above process maximizes with three electrons in the xz/yz complex
in average.)
Nonetheless, the short-range orbital correlation should persist deep into
the underdoped superconducting regime.
In addition, the light mass of orbiton and its strong coupling to the
magnon make orbiton another interesting participant in the pairing
mechanism. Furthermore, the correlated nature suggested by our study
indicates a much stronger local electron-boson coupling than the
current mean-field estimation~\cite{Boeri}. Finally, our correlated
picture also supports a strong coupling nature of superconductivity
(with strong phase fluctuation) in the underdoped regime, in
agreement with the recent observation of low super-fluid
density~\cite{Luetkens} with relatively high $T_c$.

In summary, with our first-principles Wannier function analysis, a
rare ferro-orbital order is identified in LaOFeAs that breaks the rotational
symmetry and gives rise to
the recently observed giant anisotropy of magnetic coupling in
undoped iron pnictides. The observed magnetic and lattice structures
result \textit{locally}---without widely applied frustration or the need
for Fermi surface nesting. In great contrast to the cuprates, our study
reveals the essential roles of the orbital degree of freedom in iron
pnictides in general, and suggests active roles of light orbiton in
magnon decay and electron-pairing. Finally, the pure electronic
origin and the obtained large coupling strength advocate a
correlated nature of undoped iron pnictides, and support the notion
of strong coupling superconductivity in the underdoped regime.

This work was supported by the U.S. Department of Energy,
Office of Basic Energy Science, under Contract No. DE-AC02-98CH10886,
and DOE-CMSN.

\bibliography{refs}
\end{document}